\documentclass[a4paper, aps, prd, longbibliography, preprint, superscriptaddress,nofootinbib]{revtex4-1}
\usepackage{array}
\usepackage{amsmath,amssymb,amsfonts}
\usepackage[colorlinks]{hyperref}
\usepackage{multirow}
\usepackage{float} 
\usepackage[utf8]{inputenc}
\usepackage{graphicx}
\usepackage{subfigure}
\renewcommand{\eqref}[1]{Eq.~(\textcolor{red}{\ref{#1}}) }

\def\ie{{\it i.e.}}

\def\etc{{\it etc.\ }}

\newcommand{\bew}{\begin{widetext}}
\newcommand{\enw}{\end{widetext}}
\newcommand{\bee}{\begin{equation}}
\newcommand{\ene}{\end{equation}}
\newcommand{\bea}{\begin{eqnarray}}
\newcommand{\ena}{\end{eqnarray}}
\newcommand{\bes}{\begin{subequations}}
\newcommand{\ens}{\end{subequations}}

\def\to{\rightarrow}

\def\calb{\mathcal{B}}

\def\call{\mathcal{L}}

\def\calo{\mathcal{O}}

\def\gev{\,{\rm GeV}}
\def\tev{\,{\rm TeV}}

\def\fb{\,{\rm fb}}

\def\ab{\,{\rm ab}}

\begin{document}
\title{Mono-$\gamma$ Production of a Vector Dark Matter at Future $e^+e^-$ Collider}
%

\author{Kai Ma}
\email[Electronic address: ]{makai@ucas.ac.cn}
\affiliation{School of Fundamental Physics and Mathematical Science, Hangzhou Institute for Advanced Study, UCAS, Hangzhou 310024, Zhejiang, China}
\affiliation{University of Chinese Academy of Sciences (UCAS), Beijing 100049, China}
\affiliation{Department of Physics, Shaanxi University of Technology, Hanzhong 723000, Shaanxi, China}
%
%
\date{\today}
\begin{abstract}
Associated production of a dark particle and a photon, 
represented as a mono-$\gamma$ event,
is a promising channel to probe particle contents and dynamics in the dark sector.
In this paper we study properties of the mono-$\gamma$ production of a vector
dark matter at future $e^+e^-$ colliders.
The photon-like and Pauli operators, as well as triple gauge bosons interactions
involving the dark matter, are considered in the framework of Effective Field Theory.
We show that, comparing to the Pauli operator, 
the triple gauge bosons couplings are much more interesting at high energy
collider. Beam polarization effects are also analyzed, and we show that 
the experimental sensitivities can not be enhanced significantly because of
the smaller luminosity.
\end{abstract}
\maketitle
\tableofcontents
\newpage

%
\section{Introduction}
\label{sec:intro}
Gravitational effects of dark matter (DM) have been unambiguously observed in 
astrophysical and cosmological 
measurements~\cite{Bertone:2004pz,Planck:2018vyg,Green:2021jrr}. 
Not only that, DM is also an excellent candidate of explaining some fundamental
theoretical questions of the Standard Model (SM). There are also abnormalities,
for instance the muon $g-2$~\cite{Capdevilla:2021kcf},
which can be accounted for by the DM.
However it is certainly true that the particles 
that have been observed can not be the DM.
Extending particle contents of the SM by adding new states
which interact weakly with the SM particles
is the most profound approach to study physical properties of the dark sector.
It is highly desirable to adopt an approach of 
Effective Field Theory (EFT) involving the DM, 
the so called DM EFT (DMEFT)~\cite{Birkedal:2004xn,Chae:2012bq,Barman:2021hhg},
to model-independently study physics of the dark sector. In general, the
DM can take into play at both tree and loop level~\cite{Crivellin:2014qxa,Hill:2014yka},
and a variety of theoretical models of DM have been proposed~\cite{Bertone:2018krk}.
Lots of experimental searches for DM in direct~\cite{Billard:2021uyg}, 
indirect~\cite{Gaskins:2016cha} and collider signatures~\cite{Boveia:2018yeb} 
have been conducted, but so far no clear evidence was reported.

In general, the DM can be scalar~\cite{DelNobile:2011uf,Duch:2014xda,Brod:2017bsw,Criado:2021trs,Aebischer:2022wnl}, 
fermion (Dirac or Majorana)~\cite{DeSimone:2013gj,Matsumoto:2016hbs,Brod:2017bsw,Matsumoto:2014rxa,Han:2017qkr,Duch:2014xda,Criado:2021trs,Aebischer:2022wnl} 
or vector state~\cite{Duch:2014xda,Criado:2021trs,Aebischer:2022wnl,Pospelov:2008zw}.
In this work we focus on a vector model of the DM, whose kinematical 
Lagrangian is given as,
\bee
\call_{X}
=
- \frac{1}{4} X^{\mu\nu}X_{\mu\nu} +  \frac{1}{2}m_{X}^2 X^{\mu} X_{\mu}\,,
\ene
where $X_\mu$ is the massive vector field with mass  $m_{X}$,
and $X_{\mu\nu}$ is its field strength. 
The Dark Photon (DP) model~\cite{Holdom:1985ag}, in which a dark vector state
interacts with the SM particles through a kinematical mixing term 
$\epsilon\, X^{\mu\nu} F_{\mu\nu}$~\cite{Holdom:1985ag,Fabbrichesi:2020wbt}, 
is a representation of this class.
The photon component of the DM can induce decay
into charged particles, 
and can be described by following effective operator,
\bee
\label{eq:Omix:D4}
\calo_{1} 
= 
e\epsilon\, \overline{\psi}\gamma^{\mu}\psi X_{\mu} \,.
\ene
In case of that the above kinematical mixing is the only
building block of the DM model, the mixing parameter $\epsilon$
must be very small (with a typical value $\epsilon \sim 10^{-10}$) 
such that the theory is consistent with experimental 
measurements~\cite{Essig:2013lka,Alexander:2016aln,Deliyergiyev:2015oxa,Ilten:2018crw,Bauer:2018onh}.
However, such strong constraint can be released if more new states
are involved, for instance the Axion-like particles~\cite{Ge:2021cjz}.

On the other hand, the DM can also couple to SM fermions at $D5$ via magnetic 
dipole interaction 
(Pauli operator)~\cite{Dobrescu:2004wz,Gabrielli:2016cut,Casarsa:2021rud},
\bee
\label{eq:Odipole}
\calo_{2} 
= 
\frac{1}{2\varLambda_2} \overline{\psi}\sigma^{\mu\nu}\psi X_{\mu\nu} \,,
\ene
where $\psi$ is a charged fermion in the SM, and $\varLambda_2$ is the energy 
scale parameter. 
The above dipole interaction can appear at one-loop level in a
UV completed model~\cite{Dobrescu:2004wz,Gabrielli:2016cut}.
Hence the energy scale $\varLambda_2$ 
is at the same order of mass of the heavy particles
running in the loop, which should be a scale of new physics (NP),
\ie, $\varLambda_2 \sim M_{NP}$. 
In contrast to the photon-like operator \eqref{eq:Omix:D4}, whose contribution to 
cross section of the mono-$\gamma$ process decreases as 
$1/s$ at high energy, the Pauli operator \eqref{eq:Odipole} 
can initiate the mono-$\gamma$ signal at a constant rate.
Therefore, on dimensional analysis, 
significance of the Pauli operator 
is considerably larger than the one of the photon-like operator
at high energy colliders, as long as the EFT description is valid.
So far, lower bounds on the energy scale $\varLambda_2$ has not been reported. 
It was shown that future muon colliders with center of mass energy (CoM)
$\sqrt{s} = 3\tev, 10\tev$ are expected to be good experiments to 
study it~\cite{Casarsa:2021rud}.

In this paper we study following triple gauge boson couplings,
\bea
\label{eq:NDMTGC}
\calo_{3}
&=&
\frac{1}{\varLambda^{2}_{3}}  
Z_{\mu\alpha} F^{\alpha\nu} X^{\mu}_{~\;\nu}\,,
\\[3mm]
\label{eq:NDMTGC:dual}
\calo_{4}
&=&
\frac{1}{\varLambda^{2}_{4}}  
Z_{\mu\alpha} F^{\alpha\nu} \widetilde{X}^{\mu}_{~\;\nu}\,,
\\[3mm]
\label{eq:CDMTGC}
\calo_{5}
&=&
\frac{1}{\varLambda^{2}_{5}}  
W_{\mu\alpha}^{+} W^{-\alpha\nu} X^{\mu}_{~\;\nu}\,,
\\[3mm]
\label{eq:CDMTGC:dual}
\calo_{6}
&=&
\frac{1}{\varLambda^{2}_{6}}  
W_{\mu\alpha}^{+} W^{-\alpha\nu} \widetilde{X}^{\mu}_{~\;\nu}\,,
\ena
where $F_{\mu\nu}$, $Z_{\mu\nu}$ and $W_{\mu\nu}^{\pm}$ are field strengths
of the photon, neutral and charged weak bosons, respectively;
$\widetilde{X}_{\mu\nu}$ is the dual field strength of the DM, and
$\varLambda_{3,4,5,6}$ are the energy scale parameters of the corresponding
interactions. Here and after we will call these operators as DM triple gauge boson
couplings (DMTGCs). Due to restriction of Bose statistics,
the above operators can not exist in DMEFT where the DM filed is included 
before breaking of the $SU_{W}(2)\times U_{Y}(1)$ gauge symmetry by the SM Higgs doublet~\cite{Aebischer:2022wnl}.
Similar situation happens in neutral sector of triple gauge boson interactions 
within the SM contents~\cite{Georgi:1991ch,Gaemers:1978hg,Hagiwara:1986vm}, 
\ie, couplings of $\gamma\gamma Z$, $\gamma ZZ$, 
$ZZZ$~\cite{Renard:1981es,Barroso:1984re}.
Therefore, signals of such interactions indicate either 
that there are somehow enhancement effects in some higher dimensional operators 
containing those vertices, or that the NP scale is not so far away from the 
EW scale, such that the DM field can be included effectively after the EW symmetry
breaking~\cite{Gounaris:1999kf}.
The recently reported anomaly in the W boson mass measurement can be an indication
of new physics near the EW scale~\cite{Yuan:2022cpw}.
On the other hand, when the DM couples to a SM current which is broken 
by the chiral anomaly,
Wess-Zumino type interactions between $X$ and the SM gauge bosons
can appear when heavy fermions are introduced to cancel the 
anomaly~\cite{Dror:2017ehi}.
Furthermore, in case of that $X$ couples to purely right-handed 
currents~\cite{Batell:2011qq}, the triple interaction $XWW$ will disappear,
and only the coupling $XZ\gamma$ is allowed.
Naively, experimental constraints on the charged sector of the DMTGCs 
are expected to be stronger,
and can be further complicated if kinematical mixing between the DM and 
the photon is not neglected. 
In consideration of this, we will study this part elsewhere.

In this paper, we focus on the neutral sector of the DMTGCs.
We will study the mono-$\gamma$ production of $X$ at future $e^+e^-$ colliders,
and our analysis includes photon-like and Pauli operators, as well as the DMTGCs. 
In general the vector $X$ can be always 
invisible at the colliders by assuming that it couples dominantly
to a completely dark sector,
even through its interaction with the SM particles are non-zero.
In Sec.~\ref{sec:invis}, we study invisibility of the vector $X$ by assuming
that the NP operators are the only available couplings bellow the scale
under consideration, \ie, $\varLambda_i$ (or $M_{NP}$).
This condition can give constraints on the scale parameters if 
the vector $X$ is required to be invisible at the detector.
In Sec.~\ref{sec:monap}, we study properties of the mono-$\gamma$ events
at $e^+e^-$ colliders,
including its differential cross sections (in Sec.~\ref{sec:MA:DXS}), 
and beam polarization effects (in Sec.~\ref{sec:MA:BPE}).
In Sec.~\ref{sec:limits}, we study constraints on the scale parameters 
by the BaBar experiment (in Sec.~\ref{sec:exp:BaBar}), 
DELPHE experiment (in Sec.~\ref{sec:exp:lep-delphi})
and anomalous magnetic dipole moments of the 
electron (in Sec.~\ref{sec:exp:amdm}).
Experimental significances at future colliders, CECP and ILC,
are studied in Sec.~\ref{sec:MA:ES}. Conclusions are given 
in the final section, Sec.~\ref{sec:conclusion}.

\section{Invisibility at $e^{+}e^{-}$ Collider}
\label{sec:invis}
In general, final state configuration for probing the DM at colliders 
depends on its invisibility, or its decay width $\varGamma_{X}$.
If the DM decay with a relatively high rate, then it can be visible at the detector.
The typical decay length of the $X$ in mono-$\gamma$ process at a CoM energy
$\sqrt{s}$ is given as,
\bee
L_{X} 
= \gamma_{X} \tau_{X} 
= \frac{ \sqrt{s} }{2 m_X \varGamma_{X} }  \Big( 1 + \frac{ m_X^2 }{ s } \Big)\,.
\ene
In case of that $m_X > 2 m_{\ell}$ ($\ell = e, \mu, \tau$), 
the DM can decay into charged lepton pair
via the photon-like and Pauli operators. The corresponding decay widths are given as,
\bea
\varGamma_{1}(X\to \ell^+\ell^-)
&=&
\frac{ m_X e^2 \epsilon^2 }{12\pi}
\big( 1 + 2 r_{\ell}^{2} \big) \sqrt{1 - 4 r_{\ell}^{2} } \,,
\\[3mm]
\varGamma_{2}(X\to \ell^+\ell^-)
&=&
\frac{ m_X^{3} }{24\pi\varLambda^2_2}
\big( 1 + 8 r_{\ell}^{2} \big) \sqrt{1 - 4 r_{\ell}^{2} } \,,
\ena
where $r_{\ell} = m_{\ell}/m_{X}$. The rates of these two channels are roughly at
the same order if $e\epsilon \sim m_{X}/\varLambda_{2}$.
Since the DMTGC operators, anomalous decay, $Z\to X\gamma$,
can happen when $m_{X} < m_{Z}$, 
and the corresponding decay width is given as,
\bee
\varGamma_{3(4)}(Z\to X \gamma )
=
\frac{ m_Z^{5} }{144\pi\varLambda^4_{3(4)}}
\big( 1 + r_{X}^{2} \big) \big( 1 - r_{X}^{2} \big)^3 \,,
\ene
where $r_{X} = m_{X}/m_{Z}$.
Neglecting the 3-body decay channel, 
$X\to Z^\ast\gamma \to f \overline{f}\gamma$,
then $X$ decay invisibly.
The L3~\cite{L3:1997exg} and DELPHI~\cite{DELPHI:1996qcc} collaborations 
at the LEP experiment have searched for single photons at the $Z$ resonance.
and obtain an upper limit on the branching ratio, 
$\calb_{Z\to X\gamma} < 10^{-6}$. This bound can exclude a parameter
space in the $\varLambda_{3(4)}-m_X$ plane.
We will discuss this in details in Sec.~\ref{sec:MA:ES}.

On the other hand, the DMTGC operators can initiate 3-body decays
at elementary particle level, $X \to Z^{\ast} \gamma \to f\bar{f}\gamma$,
with fermion $f$ being leptons, neutrinos or quarks. 
Such processes are suppressed by $m_{X}^{4}/m_{Z}^{4}$, 
hence it is much smaller unless $m_{X}$ is very closer to $m_{Z}$.
In case of $f=q$, the 2-body hadronic decay channels $X \to h\gamma$ 
can give non-trivial contribution. We will study these hadronic decay channels
in a separate work. For this moment, let us focus on the on-shell 
two-body decay $X\to Z \gamma$, 
which pops up if $m_{X} > m_{Z}$. 
The corresponding decay width is given as,
\bee
\varGamma_{3(4)}(X\to Z \gamma )
=
\frac{ m_X^{5} }{144\pi\varLambda^4_{3(4)}}
\big( 1 + r_{Z}^{2} \big) \big( 1 - r_{Z}^{2} \big)^3 \,,
\ene
where $r_{Z} = m_{Z}/m_{X}$. The width is suppressed by a factor of
$m_{X}^{4}/\varLambda_{3,4}^{4}$.

When we study experimental significance in Sec.~\ref{sec:MA:ES},
we will implement all the above results by requiring its decay length
is smaller than a typical value,  say $L_{X} < 1 {\rm m}$ or so,
depending on real configuration of the detector.

\section{Mono-$\gamma$ Production}
\label{sec:monap}
\begin{figure}[ht]
\begin{center}
\subfigure[]{
\label{fig:Tanfeyn}
{\includegraphics[scale=0.718]{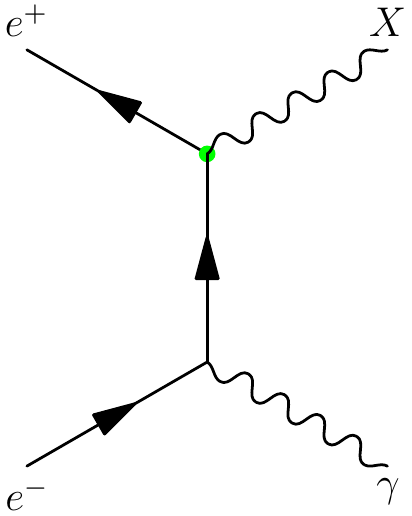}}
}
\quad
\subfigure[]{
\label{fig:Canfeyn}
{\includegraphics[scale=0.718]{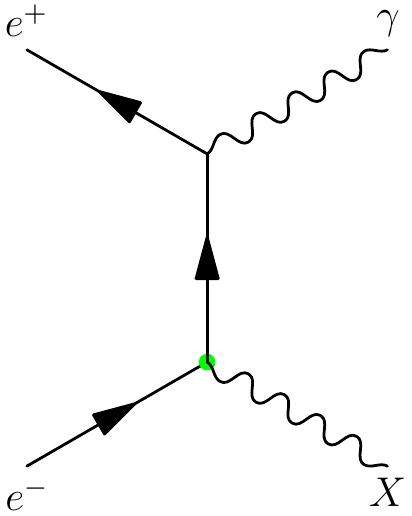}}
}
\quad
\subfigure[]{
\label{fig:Sanfeyn}
{\includegraphics[scale=0.818]{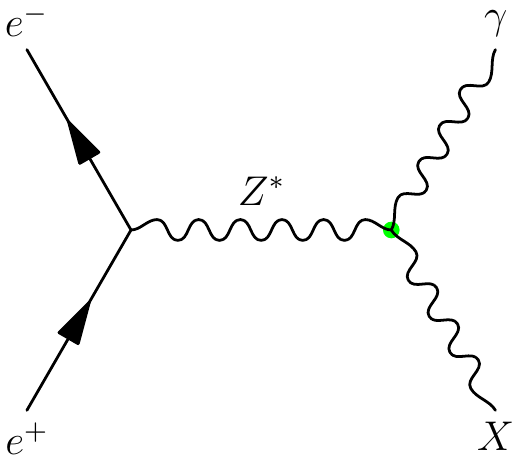}}
}
\caption{\it Feynman diagrams of the mono-$\gamma$ production 
at $e^+e^-$ collider. Fig.~\ref{fig:Tanfeyn} and Fig.~\ref{fig:Canfeyn} stand for 
initial state radiations of a photon, and can be induced by the 
operators \eqref{eq:Omix:D4} and \eqref{eq:Odipole}. In constrast,
the DMTGC operators, \eqref{eq:CDMTGC} and  \eqref{eq:CDMTGC:dual},
can generate a photon via exchange 
of a (virtual) $Z$-boson in $s$-channel, as shown in Fig.~\ref{fig:Sanfeyn}.
}
\label{fig:feyn}
\end{center}
\end{figure}

In this section we study the mono-$\gamma$ signals at $e^+e^-$ colliders, \ie, 
the process $e^+e^- \to \gamma X$.
The representative Feynman diagrams are shown in Fig.~\ref{fig:Tanfeyn}-\ref{fig:Sanfeyn}.
In cases of the photon-like and Pauli operators, the photon is generated
via initial state radiation, while for the DMTGC operators photon can be
produced via a $s$-channel exchange of a (virtual) $Z$-boson. 
The invisible $X$ results in a missing 
transverse energy at the detector. 
The dominant irreducible SM background is production of single photon
in association with a neutrino pair, \ie, $e^+e^- \to \gamma \nu\bar{\nu}$.
However, different from pair production of fermionic DM
(or production of a single fermionic DM in company with a neutrino), where
invariant mass of the missing momentum has a continuous distribution,
it is peaked at $m_{X}$ in our case because there is only one invisible particle.
This can reduce significantly the background, and the efficiency depends on
precisions of momentum measurement of the $\gamma$, beam energy 
spectrum, \etc. We will consider such effects in Sec.~\ref{sec:MA:ES}, where
experimental significances of the signals are studied.
On the other hand, one of the advantages of the $e^+e^-$ collider is 
that the beam particles can be polarized.
So, before analyzing experimental sensitivities, 
let us discuss production properties of the signals,
including differential cross section given in Sec.~\ref{sec:MA:DXS}, 
and beam polarization effects given in Sec.~\ref{sec:MA:BPE}.

\subsection{Differential and Total Cross Sections}
\label{sec:MA:DXS}
In this section we study total and differential cross section of the signals.
In massless limit of the beam particles, the polarized differential cross
sections are given as,
\bea
\label{eq:MA:DXS:MO}
\frac{d\sigma_{1,\pm\mp}}{d\cos\theta_\gamma}
&=&
\frac{ e^{2}\epsilon^2 }{4\pi s (1 - z_{X}^2)\sin^2\theta_\gamma  }
\Big[ \big( 1 + z_{X}^4 \big) \big( 1 + \cos^2\theta_\gamma \big)  
+ 2 z_{X}^2\sin\theta_\gamma^2 \Big] \,,
\\[3mm]
\label{eq:MA:DXS:PO}
\frac{d\sigma_{2,\pm\pm}}{d\cos\theta_\gamma}
&=&
\frac{ 1 }{\pi \varLambda^2_{2} (1 - z_{X}^2)  }
\Big[ \big(1 - z_{X}^2 \big)^2 + 2z_{X}^4  
+ z_{X}^2 \big(1 + z_{X}^4 \big)\cot\theta_\gamma^2
 \Big] \,,
\\[3mm]
\label{eq:MA:DXS:DMTGC}
\frac{d\sigma_{3(4),\pm\mp}}{d\cos\theta_\gamma}
&=&
\frac{ e^{2} (g_V \mp g_A)^2 \,s\, ( 1 - z_{X}^2 )^3   }
{ 64 \pi \varLambda^4_{3(4)} \big[ (1 - z_{Z}^2)^2 + z_{Z}^4 y_Z^2 \big]  }
\Big[ \sin^2\theta_\gamma + \frac{1}{2}z_{Z}^2\big( 1 + \cos^2\theta_\gamma  \big) \Big] \,,
\ena
where $z_X = m_X/\sqrt{s}$, $z_Z = m_Z/\sqrt{s}$ and 
$y_Z = \varGamma_Z/m_{Z}$; $\theta_\gamma$ is polar angle of the photon
in the Lab. frame where the $z$-axis is defined along to flying direction of the
incoming electron; 
$\sigma_{i,\, \lambda_{e^{-}}  \lambda_{e^{+}} }$ are cross section
with helicities  $\lambda_{e^{-}},  \lambda_{e^{+}} = \pm 1$ for electron and positron
, respectively.
Thanks to spin conservation, the photon-like and DTGC operators can give 
non-zero contributions only when $\lambda_{e^{-}} = - \lambda_{e^{+}}$,
while for the Pauli operator only helicity combinations with 
$\lambda_{e^{-}} = \lambda_{e^{+}}$ survive.
In addition, since parity violation of the electroweak neutral current in the SM,
$\sigma_{+-}$ and $\sigma_{-+}$ has a difference depending on 
product between vector and axial-vector couplings, \ie, $g_{V}g_{A}$ of the
electron (see \eqref{eq:MA:DXS:DMTGC}).

\begin{figure}[t]
\centering
\subfigure[]{
\label{fig:DXS:cosa}
\includegraphics[width=0.46\textwidth]{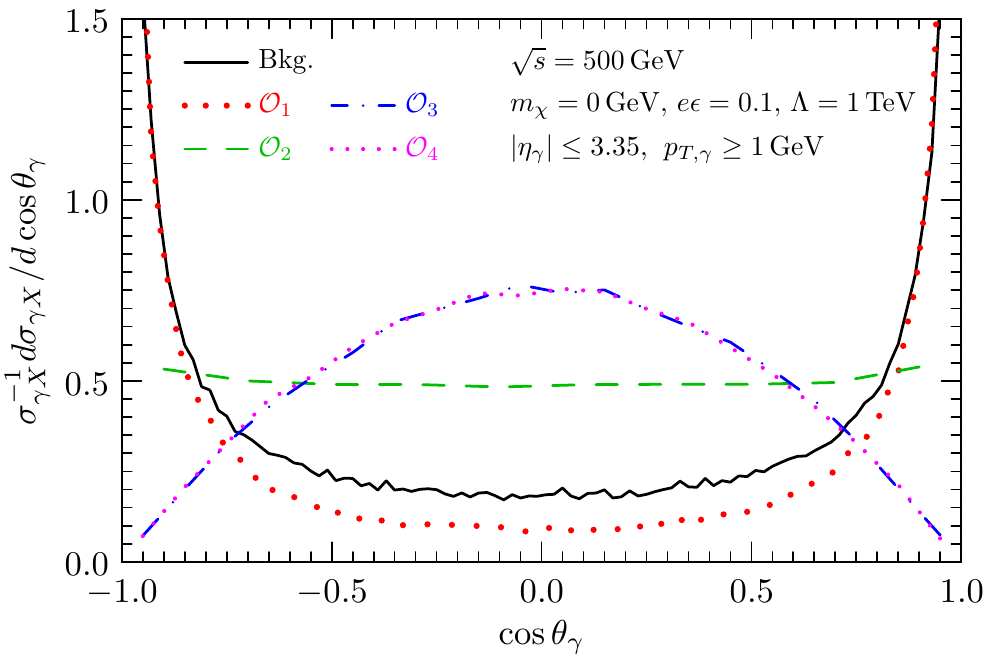}
}
\subfigure[]{
\label{fig:DXS:ea}
\includegraphics[width=0.46\textwidth]{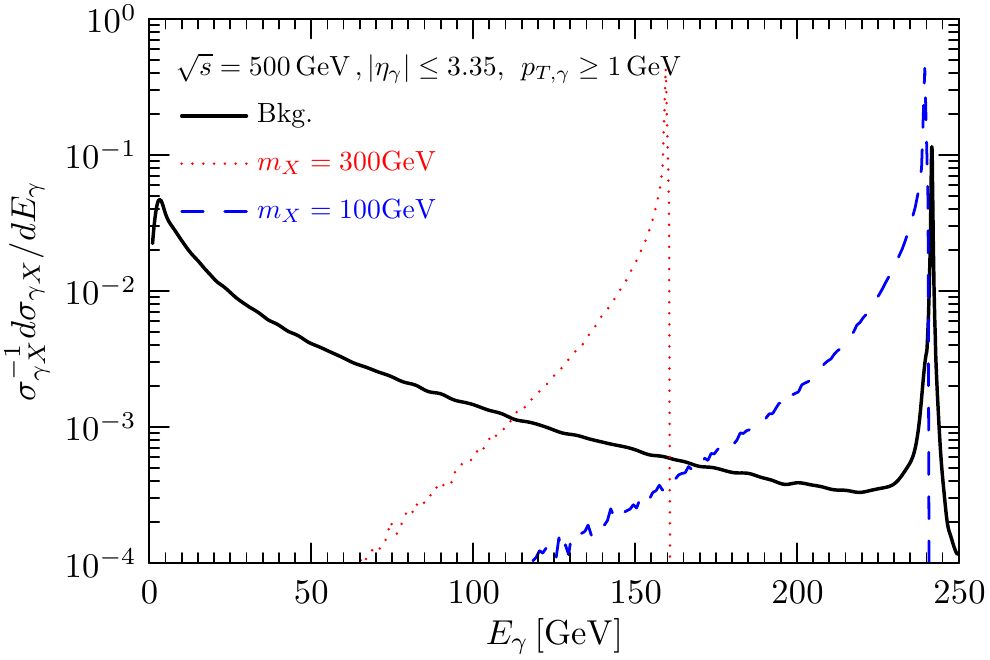}
}
\caption{\it 
Fig.~\ref{fig:DXS:cosa}: normalized distributions of the polar angle of the 
photon in the CoM frame with $\sqrt{s} = 500\gev$ and $m_{\chi}=0\gev$.
Fig.~\ref{fig:DXS:ea}: normalized distributions of the energy of the photon 
in the CoM frame with $\sqrt{s} = 500\gev$. 
Since $E_\gamma$ is determined by purely kinematical
condition, distributions of $E_\gamma$ for the signals are illustrated by the
DMTGC operators with $m_{\chi}=100\gev$ and $m_{\chi}=300\gev$.
In addition we have put kinematical cuts $ \big|\eta_\gamma \big| < 3.35$
and $p_{T,\gamma} > 1\gev$ in both plots.
}
\label{fig:xca-ea:yxs}
\end{figure}

We can also see that there are singularities at $\theta_\gamma=0,\pi$
for the photon-like operator, similar to the well-known property of the background.
Such singularity disappears in channels induced by the Pauli and DMTGC operators.
In order to avoid those kinematical space, we implement our operators into
\texttt{FeynRules}~\cite{Alloul:2013bka}, 
and using \texttt{MadGraph5}~\cite{Alwall:2014hca}
to estimate cross sections with following kinematical cuts,
\bee
\big| \eta_{\gamma} \big| \le 3.35,\quad
p_{T,\gamma} > 1\gev\,.
\ene
Fig.~\ref{fig:DXS:cosa} shows polar angle distributions of the signal
and background with above kinematical cuts 
at a typical center of mass energy $\sqrt{s}= 500\gev$,
and for the signal we have set $m_{X}=0$, $e\epsilon=0.1$, 
and $\varLambda_i=1\tev$ for reference. 
Signal of the photon-like operator is much similar to the background.
However, the Pauli operator possesses a constant polar angle distribution,
as can be seen from \eqref{eq:MA:DXS:PO} in case of $z_{X}=0$.
For the DMTGC operators, the two operators have completely
the same distribution, as we can seen from \eqref{eq:MA:DXS:DMTGC}.
However, signal events are dominated in the central region of the detector.
This is completely different from the background, 
and results in a higher kinematical selection efficiency.
Furthermore, this property is purely because of transverse part contribution
of a spin-1 particle exchanged in $s$-channel, 
and hence is independent on mass of the DM, 
$m_{X}$ (see \eqref{eq:MA:DXS:DMTGC}).
The above polar angle distributions can be used to distinguish the
new physics operators, and we will discuss the detials in Sec.~\ref{sec:MA:ES}.

Neglecting practical limitations on experimental measurements,
energy of the radiated photon has a fixed value for the signal. 
For CoM energy $\sqrt{s}$ it is given as,
\bee
\label{eq:e:monoa}
E_\gamma = \frac{1}{2} \sqrt{s} \left( 1 - \frac{m_{X}^{2} }{s} \right)\,.
\ene
This is determined by purely kinematical conditions, the photon
has the same energy $E_\gamma$ for all operators.
However, photons of the background
have a continuous energy spectrum dominated at soft region,
as shown in Fig~\ref{fig:DXS:ea}.
The peak at $ ( 1 - m_{Z}^{2} /s )\sqrt{s}/2 \approx 241.7\gev$ 
is due to the resonant channel
$e^+e^- \to Z(\nu\bar{\nu})\gamma$. This can introduce a problem 
for probing signals when $m_{X}$ is near $m_{Z}$.
The situation gets worse if $\sqrt{s} \gg m_{Z}, m_{X}$, 
in which case $E_\gamma \sim \sqrt{s}/2$ for both signals and background.
In any case, since energy of the signal is always peaked at $E_\gamma$,
background events can be suppressed by a factor of $10^{-1}$ to $10^{-4}$,
depending on mass of the DM.
Nevertheless, Due to Initial State Radiation (ISR) and 
emission of beamstrahlung photons~\cite{Yokoya:1985xx},
beam energies are characterized by continuous spectra.
Energy of the photon is hence smeared.
In Fig~\ref{fig:DXS:ea}, distributions of $E_\gamma$ for $\calo_3$
are shown for $m_X=100\gev$ and $300\gev$.
The ISR effect is taken into account by using the plugin 
\texttt{MGISR}~\cite{Chen:2017ipx,Li:2018qnh} to the \texttt{MadGraph}.
We can see clearly the smearing effect. It turns out that 
selection efficiency of the signal is reduced.

\begin{figure}[t]
\begin{center}
\subfigure[]{
\label{fig:MA:XE-YXS}
\includegraphics[width=0.46\textwidth]{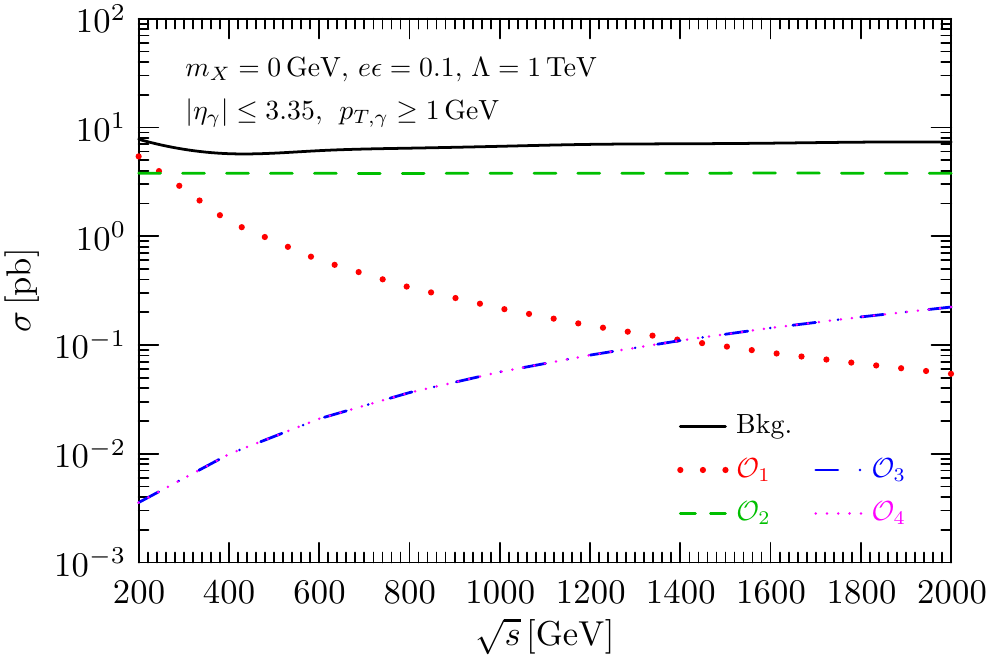}
}
\subfigure[]{
\label{fig:MA:MX-YXS}
\includegraphics[width=0.46\textwidth]{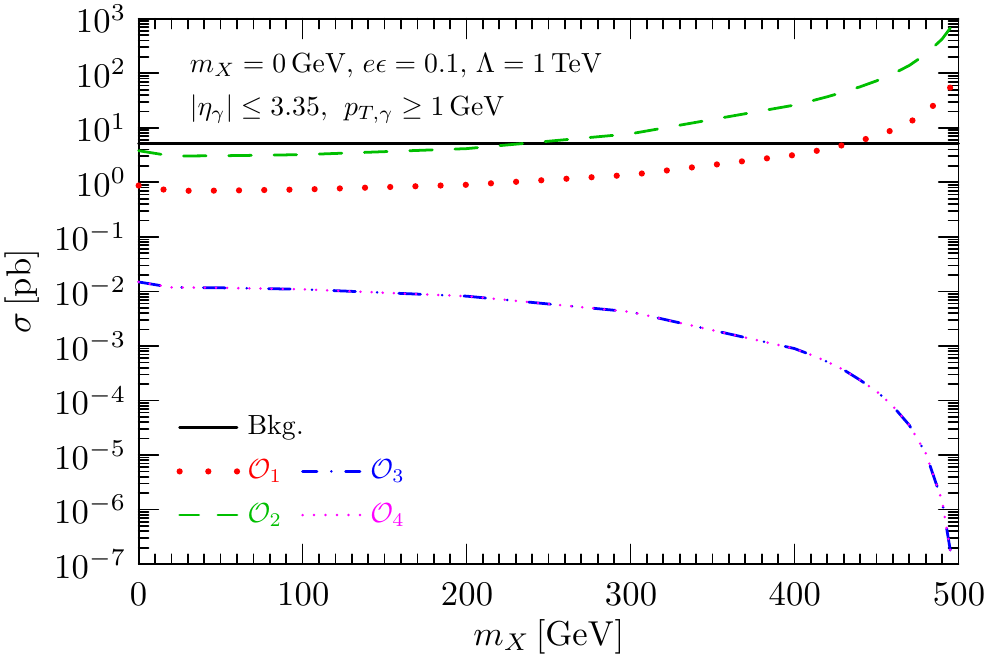}
}
\caption{\it 
Fig.~\ref{fig:MA:XE-YXS}, 
cross sections of mono-photon productions with respect to center of mass energy
$\sqrt{s}$ for $m_{\chi}=0\gev$, $e\epsilon=0.1$ and $\varLambda_i=1\tev$
Fig.~\ref{fig:MA:MX-YXS}, mass dependences of the total cross section 
at $\sqrt{s}=500\gev$ for  $e\epsilon=0.1$ and $\varLambda_i=1\tev$.
In addition we have put kinematical cuts $ \big|\eta_\gamma \big| < 3.35$
and $p_{T,\gamma} > 1\gev$ in both plots.
}
\end{center}
\end{figure}
Fig.~\ref{fig:MA:XE-YXS} and \ref{fig:MA:MX-YXS} show CoM
energy and mass dependences of the total cross section, respectively.
we can see that, while cross section of the photon-like 
operator is dominant at low energy,
and the one of the Pauli operator keeps a constant in the whole range of $\sqrt{s}$, 
contributions of the DMTGC operators grow quickly with in creasing CoM energy.
On the other hand, cross section of the background is reduced slightly
from $\sqrt{s}=200\gev$ to $\sqrt{s}=400\gev$. 
This behavior closely related to the resonance channel
$e^+e^- \to Z(\nu\bar{\nu})\gamma$.
The cross section reachs roughly a constant at high energy region.
For the DMTGC operators, the $m_{X}$ dependence shows a normal
kinematical suppression at large mass region. However, 
for both the photon-like and Pauli operators 
the distributions show an enhancement as $m_X \to \sqrt{s}$.
This is due to soft singularity (in massless limit of the incoming electrons),
as can be seen in \eqref{eq:MA:DXS:MO} and \eqref{eq:MA:DXS:PO} 
(there is a factor $1-z_{X}^{2}$ in the denominator in both cases).

\subsection{Beam Polarization Effects}
\label{sec:MA:BPE}
One of the most advantages of the $e^+e^-$ collider is that the beam particles 
can be polarized. Since the background contributes mainly through chiral
couplings of the SM, particularly the $e\nu W$ coupling at high energy region,
polarized beams are much help to reduce the background.
The cross section with electron beam polarization $P_{e^-}$
and positron beam polarization $P_{e^+}$ are given by,
\bea
\sigma\big(P_{e^-}, P_{e^+}\big)
&=&
\frac{1}{4} \sum_{ \lambda_{e^{-}} , \lambda_{e^{+}} = \pm 1}
\big( 1 + \lambda_{e^{-}} P_{e^-} \big) \big( 1 + \lambda_{e^{+}} P_{e^+} \big) 
\sigma_{\lambda_{e^{-}}  \lambda_{e^{+}} }  \,,
\ena
where $\sigma_{\lambda_{e^{-}}  \lambda_{e^{+}} }$ are cross sections
with 100\% polarizations,
and for signals they are given in \eqref{eq:MA:DXS:MO}-\eqref{eq:MA:DXS:DMTGC}.
\begin{figure}[t]
\centering
\includegraphics[width=0.48\textwidth]{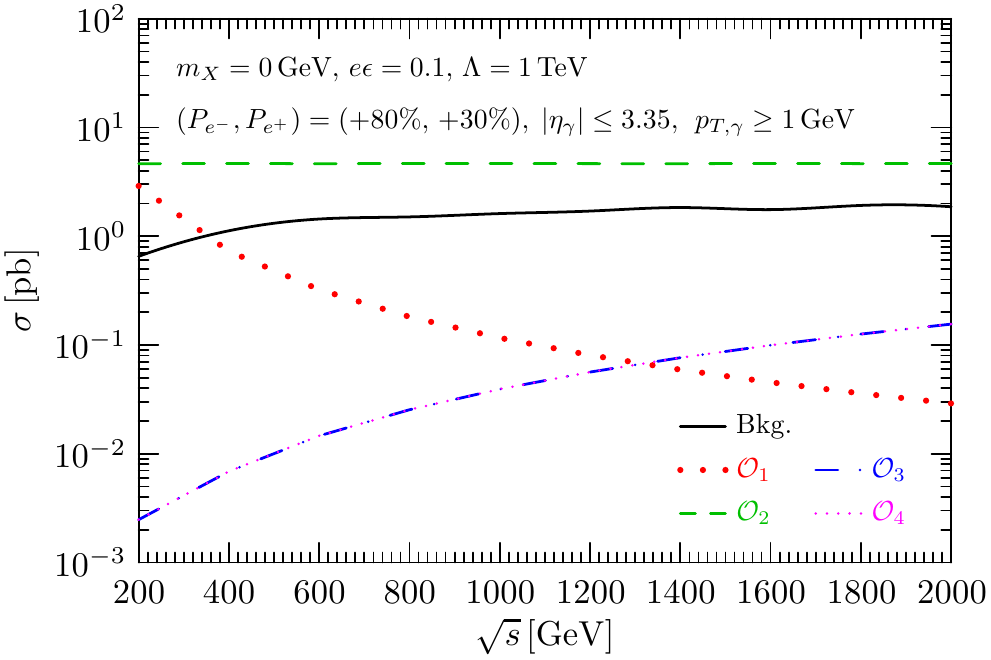}
\quad
\includegraphics[width=0.48\textwidth]{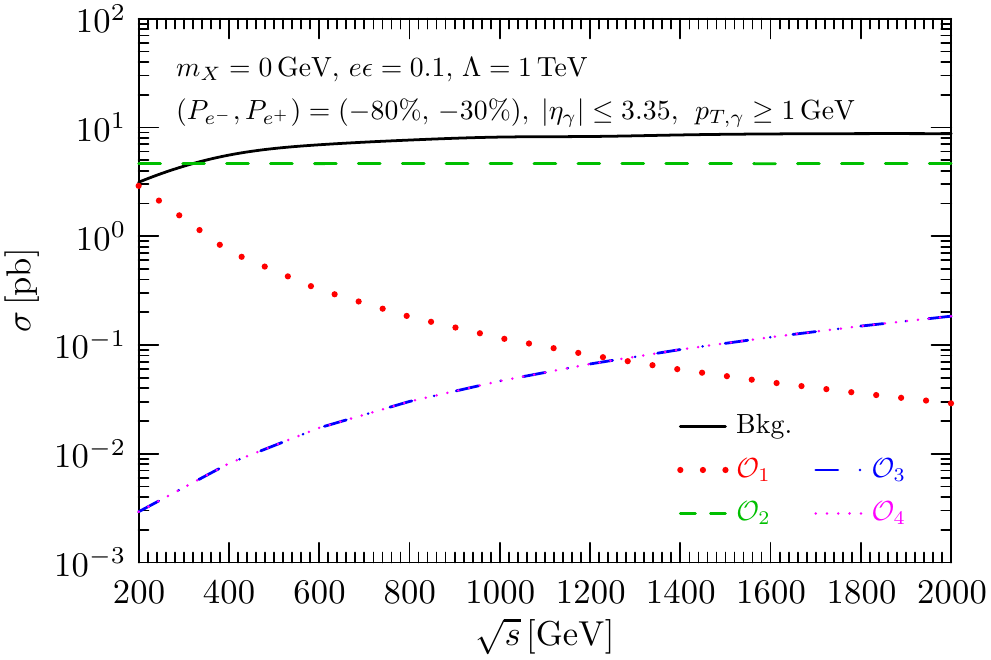}
\\
\includegraphics[width=0.48\textwidth]{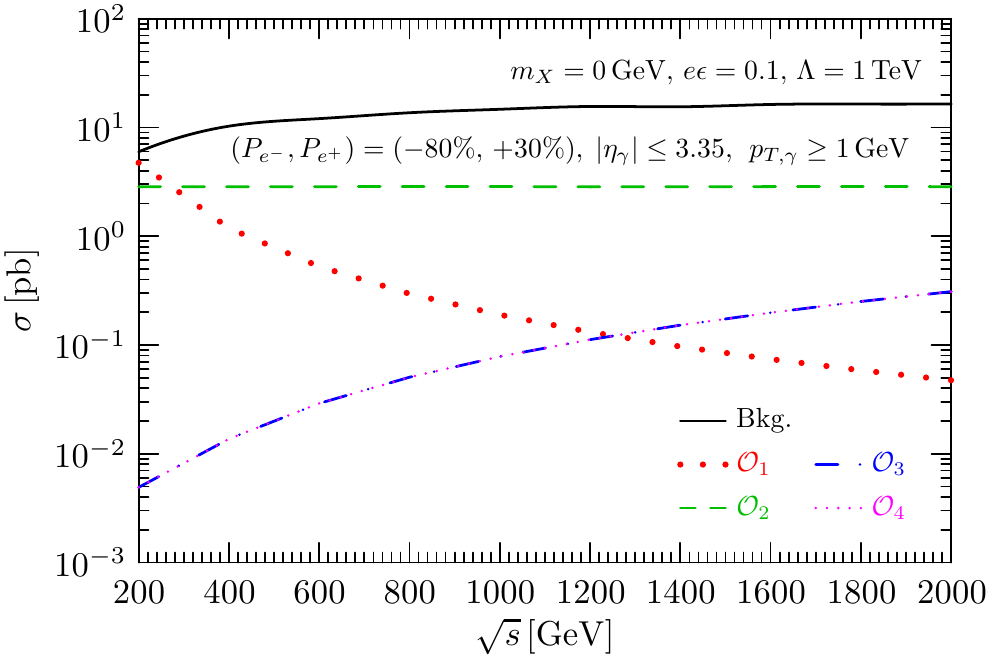}
\quad
\includegraphics[width=0.48\textwidth]{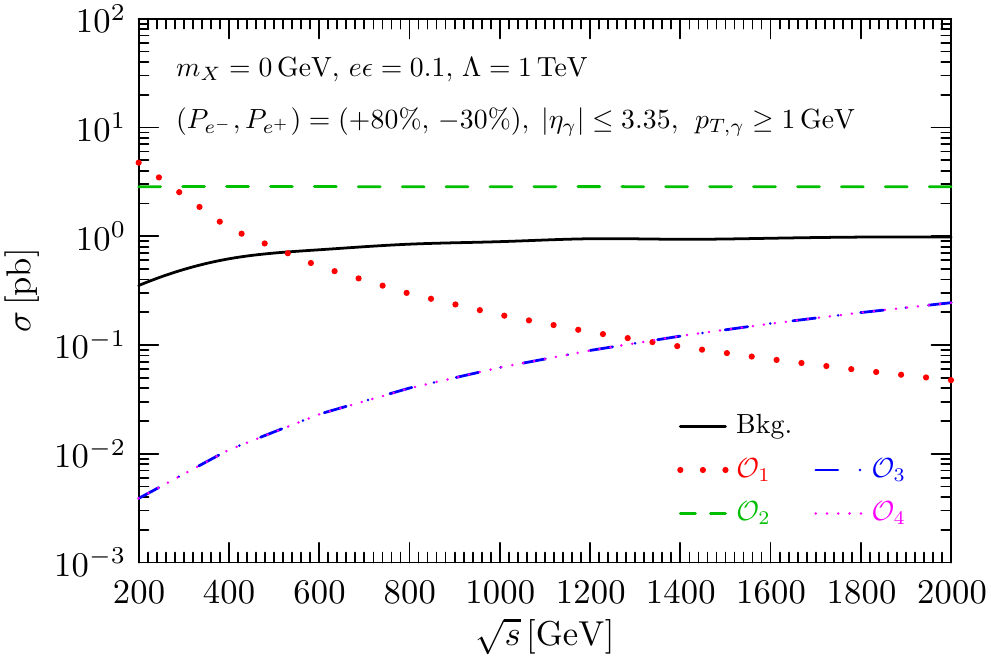}
\caption{\it 
Energy dependences of the cross sections with beam polarizations
$(P_{e^-},\, P_{e^+})=(+80\%, -30\%)$ (top-left panel),
$(-80\%, -30\%)$ (top-right panel),
$(-80\%, +30\%)$ (bottom-left panel),
and $(+80\%, -30\%)$ (bottom-right panel), respectively.
Signals are shown with $m_{X}=0\gev$, $e\epsilon=0.1$ and $\varLambda_i=1\tev$. 
}
\label{fig:xe-yxs:pol}
\end{figure}
Fig.~\ref{fig:xe-yxs:pol} shows the polarized cross sections with typical polarization
$P_{e^-}=\pm80\%$ and $P_{e^+}=\pm30\%$.
For the background, the cross sections are shown without contributions
of the resonance channel ($e^+e^- \to Z(\nu\bar{\nu})\gamma$), which
is less affected by beam polarizations, and is inessential when
$E_\gamma$ is not closer to $E_\gamma^{Z}$.
The rest contributions come from the left-handed charged currents in the SM,
the dominant background is $\sigma_{-+}$.
Hence $\sigma_{\rm Bkg}(-80\%, +30\%)$ is the largest one, 
as shown in the bottom-left panel. The other polarized channels
are roughly suppressed by a factor from $0.06$ to $0.54$.
For the Pauli operator, chirality is fliped in the neutral current, 
$\sigma_{\pm\pm}$ are the only non-zero contributions.
Hence the largest polarized channels are $\sigma_{2}(\pm80\%, \pm30\%)$,
the others are suppressed by a factor of about $0.61$.
In contrast, for both photon-like and DMTGC operators, 
the non-vanishing $100\%$ polarized cross sections are $\sigma_{\pm\mp}$. 
This property results in the largest contributions 
$\sigma_{1,3,4}(\pm80\%, \mp30\%)$.
Contributions of the other polarization configurations 
are reduced by a factor of about $0.61$.

\section{Constraints from $e^+e^-$ Experiments and $a_{e}$}
\label{sec:limits}
In this section we study constraints on the various NP operators by 
searches at the BarBar~\cite{BaBar:2017tiz}
and DELPHI~\cite{DELPHI:2008uka} experiments,
as well as measurements of the anomalous magnetic dipole moment
of the electron~\cite{Morel:2020dww,Darme:2020sjf}.
There are also astrophysical and cosmological constraints~\cite{Dror:2017ehi}, 
particularly bounds on the photon-like operator are very strong~\cite{Ge:2021cjz}, 
but we don't consider those limits here.

\subsection{BaBar}
\label{sec:exp:BaBar}
The BaBar experiment with CoM energy of $10.58\gev$
at the PEP-II B-factory has searched for dark photon
by mono-$\gamma$ events with  a total luminosity of $53\fb^{-1}$~\cite{BaBar:2017tiz}.
A search for dark photon in the resonance channel~\cite{BaBar:2014zli},
$e^+e^- \to \gamma X,\, X \to \ell^+\ell^-$ ($\ell = e,\mu$)
was also conducted by the BaBar Collaboration.
However, exclusion limits in this channel depend on branching ratio
of the decay $X \to \ell^+\ell^-$.
Here we re-interpret mono-$\gamma$ results for the Pauli and DMTGC operators.
The single photon was required to have a polar angle in the following range,
\bea
-0.4 < \cos\theta_\gamma < 0.6\, && \text{for} \quad m_X < 5.5\gev\,,
\\[2mm]
-0.6 < \cos\theta_\gamma < 0.6\, && \text{for} \quad m_X > 5.5\gev\,,
\ena
in the CoM frame. Here $m_X=5.5\gev$ is the critical value for defining 
low ($m_X < 5.5\gev$) and high mass ($m_X>5.5\gev$) regions.
In addition, photon is further selected by the cuts $E_\gamma > 3\gev$
and $E_\gamma > 1.5\gev$ in the low and high mass regions, respectively.
The cuts $E_\gamma > 3(1.5)\gev$ are helpful to reduce background,
but are useless for signals since the polar angle requirements
have rejected the events with $E_{\gamma} \lesssim 3.86\gev$ in the low mass region
and  $E_{\gamma} \lesssim 2.27\gev$ in the high mass region 
(with $m_X < 8\gev$ which corresponds to the maximum searched by the experiment).
Hence we will ignore effects of the cuts on $E_\gamma$.
On the other hand, as we have seen in Fig.~\ref{fig:DXS:cosa}
that polar angle distributions of the operators are completely different,
therefore, efficiencies of the geometric cuts can be very different.
The BaBar Collaboration used Boosted-Decision-Tree (BDT) 
based on characteristics of $E_\gamma$ and $\cos\theta_\gamma$
to select signals.
Here we consider only the effects of the geometric cuts, 
and are accounted for by implementing the cuts at generator level. 
The corresponding acceptance efficiencies are estimated at selected 
representative points 
$(\epsilon=\epsilon_{\rm BarBar}, m_{X}=m_{X}^{\rm BarBar} )$
on the 90\%C.L. exclusion line of the BaBar.
Assuming that the trigger and 
reconstruction efficiencies of the photon are the same for all the operators, 
then for given mass $m_{X} = m_{X}^{\rm BarBar}$, 
the 90\%C.L. lower limits on the energy scales $\varLambda_{i}$ are given as,
\bee
\varLambda_{i} 
\ge
\left[ 
\frac{\sigma_{i}(\varLambda_{i}=1\gev,  m_{X}=m_{X}^{\rm BarBar} )} 
{ \sigma_{1}(\epsilon=\epsilon_{\rm BarBar}, m_{X}=m_{X}^{\rm BarBar} ) }
 \right]^{1/\kappa}
 \big[ \gev \big] \,,
\ene
where $\kappa = 2, 4$ for $i=2, 3(4)$, respectively;
and the cross sections are calculated after the geometric cuts.
Our results will be shown in Sec.~\ref{sec:MA:ES}.

\subsection{LEP-DELPHI}
\label{sec:exp:lep-delphi}
Constraints on emission of an invisible graviton from low-scale extra-dimension
and supersymmetric models 
were studied by the DELPHI experiment at LEP~\cite{DELPHI:2008uka},
and has been re-interpreted as limits on Dark Matter, 
for instance, in Ref.~\cite{Fox:2011fx}.
The DELPHI data was obtained with different CoM energies~\cite{DELPHI:2003dlq}, 
ranging from 180.8\gev to 209.2\gev.
The single-photon events were selected by three different triggers:
the High density Projection Chamber (HPC), 
the Forward ElectroMagnetic Calorimeter (FEMC) 
and the Small angle TIle Calorimeter (STIC).
Here we focus on the HPC which covers a wider range of $E_\gamma$,
\bee
\label{eq:kc:delphe}
45^\circ < \theta_{\gamma} < 135^\circ\,,
\quad 
0.06 E_{\rm Beam} < E_\gamma < 1.1 E_{\rm Beam}\,,
\ene
comparing to FEMC and STIC.
However, the HPC module has relatively lower trigger efficiency and worse
energy resolution. The trigger efficiency strongly depends on the photon energy,
and is about 52\% at $E_\gamma = 6\gev$ and 
above 77\% when $E_\gamma > 30\gev$, and reached a maximum
84\% when $E_\gamma \simeq E_{\rm Beam}$. 
Since energies of the photons generated with parameters considered in this paper 
are almostly larger than $20\gev$, we will use a constant trigger efficiency  
$\epsilon^{\rm Trig} = 80\%$ in our analysis.
Furthermore, energy of the mono-$\gamma$ (given by \eqref{eq:e:monoa})
can also be significantly smeared by the energy resolution,
particularly for hard photons (or equivalently smaller $m_X$).
It turns out that selection efficiencies of the signals decrease in the low 
$m_X$ region. In our following analysis, we require 
$\left| E_{\gamma} - E_\gamma(m_X) \right| < 1\gev$,
and energy dependence of the efficiency
due to energy resolution was taken into account by
the systematical uncertainty $\pm8\%$~\cite{DELPHI:2003dlq}.
The experimental significances are estimated by calculating
following $\chi^2$ function,
\bee
\chi^2
=
\sum_{i,j}
\left[ \frac{ 
N_{\rm Sig}\big( \sqrt{s_{i}},  \overline{\theta}_j  \big)  }
{ \sqrt{  N_{\rm Bkg+Sig}\big( \sqrt{s_{i}},  \overline{\theta}_j  \big) +
\Delta\sigma_{\rm Syst}^2 \cdot
N_{\rm Bkg}^{2}\big( \sqrt{s_{i}},  \overline{\theta}_j  \big) } 
} \right]^2\,,
\ene
where $N_{\rm Bkg/Sig/Bkg+Sig}\big( \sqrt{s_{i}},  \overline{\theta}_j  \big) = \call_{i} \cdot \epsilon^{\rm Trig}  \cdot \sigma_{\rm Bkg/Sig/Bkg+Sig} \big( \sqrt{s_{i}},  \overline{\theta}_j  \big)$
are number of events of the background, signal and summation of the background
and signal, respectively; and
polar angle distribution has also been binned in order to enhance
the significance, and $\overline{\theta}_j$ stands for 
$c_{j-1} <|\cos\theta_{\gamma, j}| < c_j$ 
with $c_{j}$ being the boundary values of the bin.
The results will be discussed in Sec.~\ref{sec:MA:ES}.

\subsection{Anomalous Magnetic Dipole Moment}
\label{sec:exp:amdm}
Discrepancy between theoretical prediction and experimental measurements 
in the magnetic dipole moment of the muon has been reported long ago
(we refer to Refs.~\cite{Aoyama:2020ynm,Keshavarzi:2021eqa} for recent reviews).
Combination of the recent measurements by the FNAL Muon $g-2$ 
experiment~\cite{Muong-2:2021ojo} and  the old BNL 
result~\cite{Muong-2:2006rrc} has pushed the discrepancy to a level of $4.2\sigma$.
Implications of this anomaly have been widely studied.
In this paper we focus on the anomalous magnetic dipole moment of the electron.
Improved measurement of the fine structure constant by a matter-wave interferometer 
of cesium-133 atoms~\cite{Parker:2018vye}
shows a $2.4\sigma$ tension with the SM prediction~\cite{Aoyama:2017uqe},
\bee
\Delta a_e
\equiv
a^{\rm Exp}_e - a^{\rm SM}_e = -(8.8 \pm 3.6)\times 10^{-13}
\quad \text{(\rm Berkeley-2018)}\,.
\ene
The important thing is that it has a different sign from the $\Delta a_{\mu}$.
Even through it is still suggestive, 
such a discrepancy challenges theoretical models which try to explain 
both $\Delta a_{e}$ and $\Delta a_{\mu}$ simultaneously.
However, the most recent atomic physics measurement of $\alpha_{\rm em}$
using Rubidium-87 atoms implies~\cite{Morel:2020dww,Darme:2020sjf},
\bee
\label{eq:LKB2020}
\Delta a_e
\equiv
a^{\rm Exp}_e - a^{\rm SM}_e = +(4.8 \pm 3.0)\times 10^{-13}
\quad \text{(\rm LKB-2020)}\,,
\ene
which is more than $4\sigma$ away from the Berkeley-2018 result.
More interestingly, the deviation is positive, has the same sign with the $\Delta a_{\mu}$.
Even through experimental uncertainties are at the same level,
it is clear that further improved measurements are
necessary to clarify the discrepancies reported in these two experiments,
and consistent experimental results can be expected in forthcoming years.
In following analysis we ignore this sign problem,
and use the result given in \eqref{eq:LKB2020}
to study constraints on the mixing parameter $\epsilon$.
Contribution of the photon-like operator to $\Delta a_{e}$ is given 
as~\cite{Fayet:2007ua,Pospelov:2008zw},
\bee
\Delta a_{e}^{X}
=
\frac{ e^2 \epsilon^2 }{4\pi^2} \, r_e^2  \, F_{X}(r_e) \,,
\ene 
where $r_e = m_e/m_X$, and the function $F_{X}(r_e)$ is given as,
\bee
F_{X}(r_e)
=
\frac{1}{2} \int_0^1 dx \frac{ 2x^2(1-x) }{ (1-x)(1 - r_e^2 x) + r_e^2 x }\,,
\ene
which is always positive.

\section{Significances at Future $e^+e^-$ Colliders}
\label{sec:MA:ES}
In practical measurement, energy of the photon can be 
smeared by, for instance, ISR of the beams and beamstrahlung emission of the 
photons~\cite{Habermehl:2020njb}.
In order to avoid overestimation of the experimental sensitivities,
and also to select most of the signal events, the above detector
activity should be taken into account.
Take ILC as an example~\cite{Habermehl:2020njb},
it was shown that nearly 70\% of the beam particles have energy
lying in the window $\big| E_{\rm Beam} - 250\gev \big| \le 1\gev$,
see also our simulation results in Fig.~\ref{fig:DXS:ea}.
In following calculations of experimental sensitivity, 
we estimate both signal and background events 
by assuming that the beam energy has a fixed 
value $\sqrt{s}/2$, and cross section of the signal is multiplied by an efficiency factor,
$\epsilon_{\rm ISR} = 70\% $,
no matter which collider is under considering.
Furthermore, the background is estimated by collecting all the cross sections
as long as energy of the photon lies in the window 
$\big| E^{\rm Bkg}_\gamma - E^{\rm Sig}_\gamma \big| < 1\gev$.
The above simple approximation does not capture all signal 
(and also background) information, but much conservative it is.
On the other hand, as we have shown that in Fig.~\ref{fig:DXS:cosa},
polar angle of the photon is much sensitive to the signal, 
and hence useful to enhance the experimental sensitivities.
In consideration of this,
distribution of the variable $\cos\theta_{\gamma}$ is 
divided into 10 bins, and
the experimental significance is estimated by calculating 
following $\chi^2$ function,
\bee
\label{eq:chi2}
\chi^2 
= 
\sum_{i}
\frac{ \left(\epsilon_{\rm ISR} \cdot N^{\rm Sig}_i \right)^2 }
{  N^{\rm Bkg}_i  +  \epsilon_{\rm ISR} \cdot N^{\rm Sig}_i  +
\left( \epsilon_{\rm Syst} \cdot N^{\rm Bkg}_i \right)^2 } \,,
\ene
where $N^{\rm Sig}_i$ and $N^{\rm Bkg}_i$ are signal and background
events in the $i$-th bin, $\epsilon_{\rm Syst}$ stands for systematic uncertainty 
which can reduce the sensitivity significantly as shown in Ref.~\cite{Kundu:2021cmo}.
We also assume that $\epsilon_{\rm Syst} = 1\%$ in following calculations.
The radiative BhaBha process can also contribute the background.
However such contribution can be significantly reduced by requiring that there is
only one reconstructed BeamCal cluster, as reported in Ref.~\cite{Habermehl:2020njb}.
So, we will neglect the background coming from radiative BhaBha process.

\subsection{Significances at CEPC}
\label{sec:MA:ES:CEPC}
\begin{figure}[ht]
\begin{center}
\subfigure[]{
\label{fig:S:CEPC:O1}
{\includegraphics[width=0.3\linewidth]{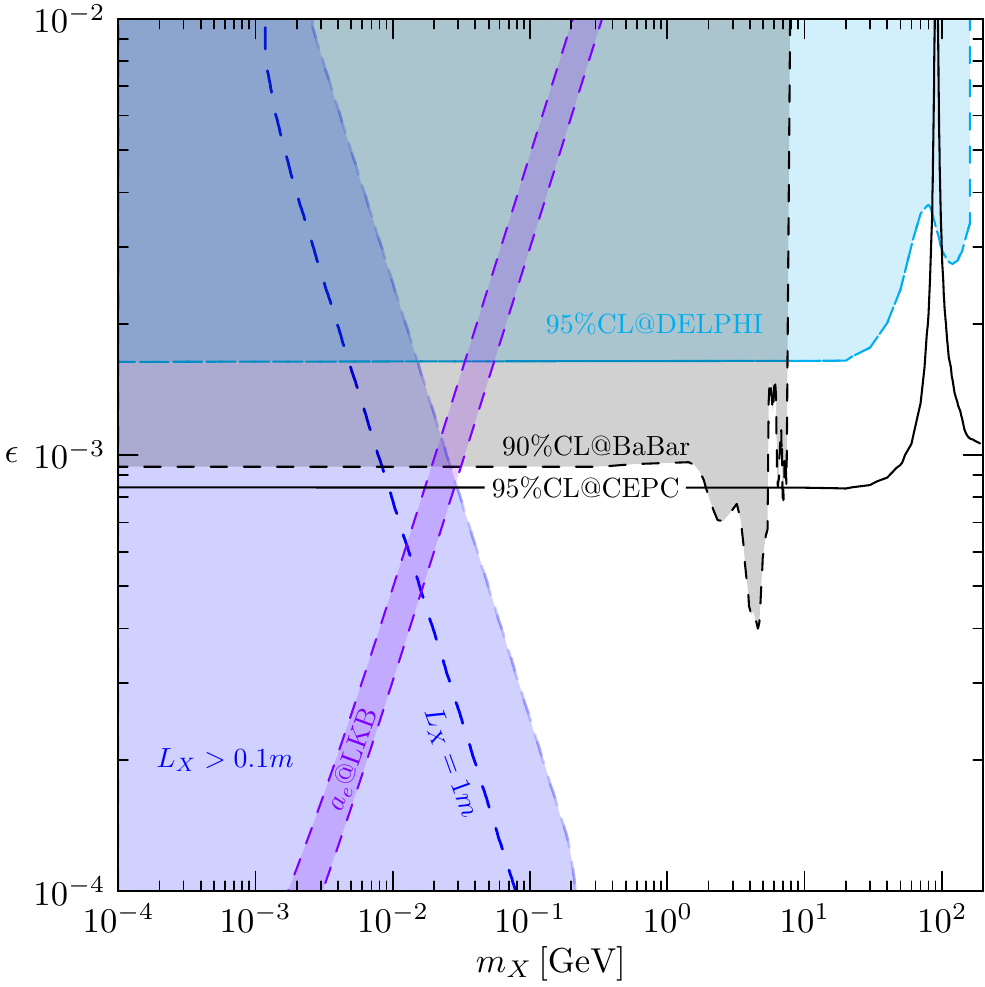}}
}
\subfigure[]{
\label{fig:S:CEPC:O2}
{\includegraphics[width=0.3\linewidth]{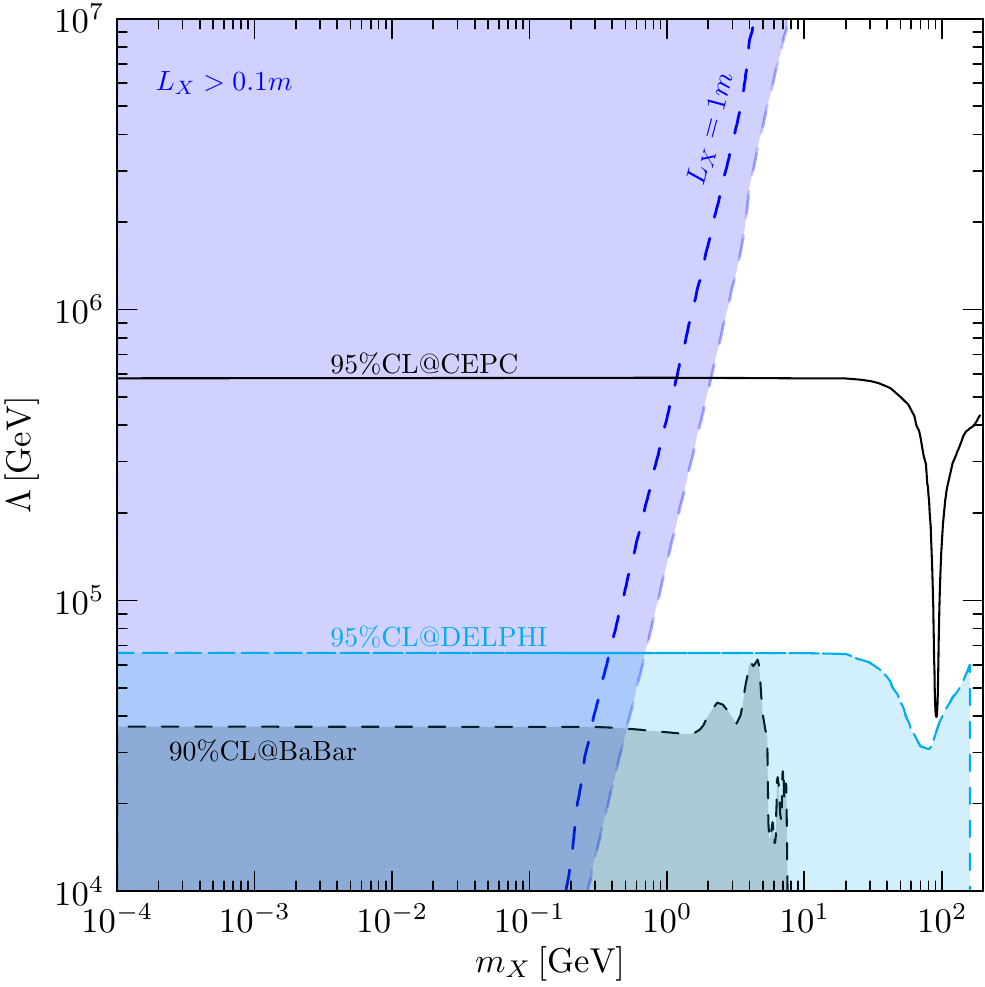}}
}
\subfigure[]{
\label{fig:S:CEPC:O3}
{\includegraphics[width=0.3\linewidth]{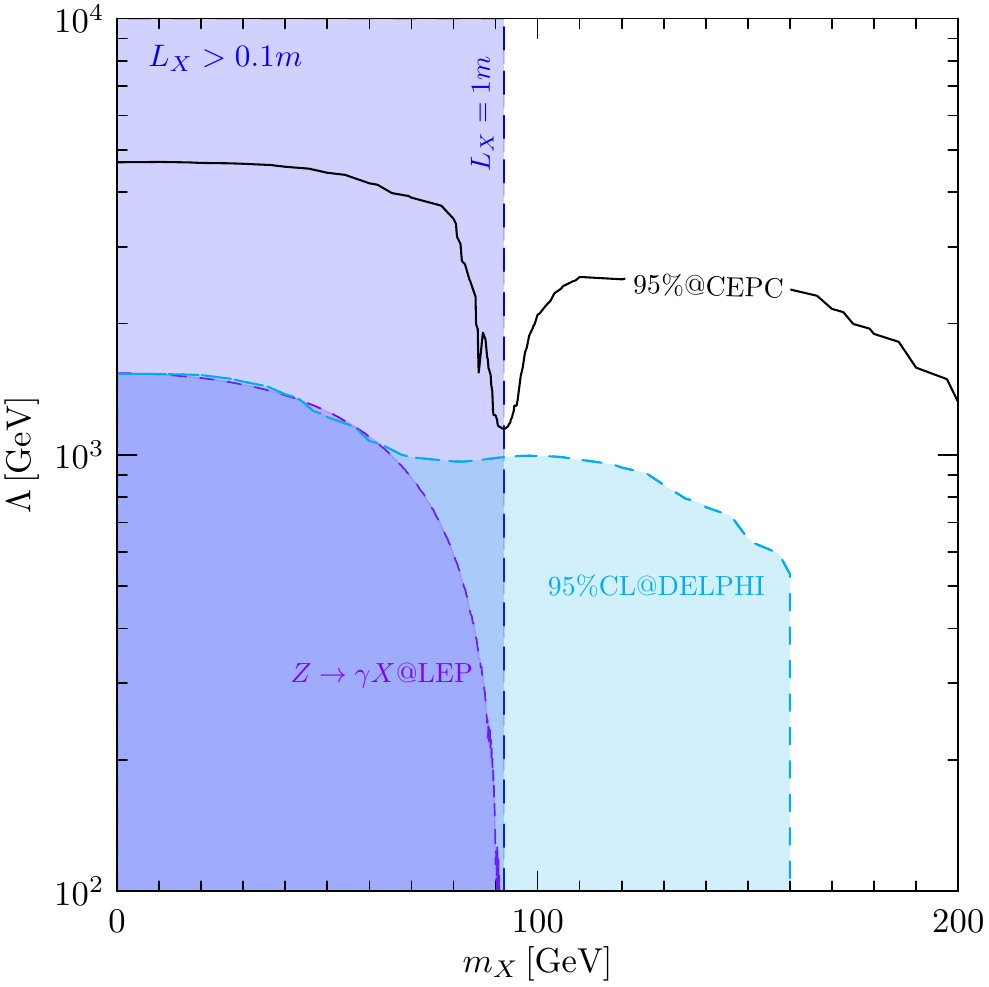}}
}
\caption{\it
Expected sensitivities at 95\%CL for the operators 
$\calo_1$ (Fig.~\ref{fig:S:CEPC:O1}), $\calo_2$(Fig.~\ref{fig:S:CEPC:O2}) 
and $\calo_{3(4)}$ (Fig.~\ref{fig:S:CEPC:O3})
at the CEPC with a total luminosity $5.6\ab^{-1}$.
The shades regions represents constraints from other experiments,
and explained in the text.
}
\label{fig:S:CEPC}
\end{center}
\end{figure}

The CEPC experiment is designed to be a Higgs factory~\cite{CEPCStudyGroup:2018ghi}, 
but is also relevant for probing particles and dynamics 
in dark sector~\cite{Liu:2019ogn}.
Three different running modes at the CEPC have been proposed~\cite{CEPCStudyGroup:2018ghi}.
In this study we focus on the mode with $\sqrt{s} = 240\gev$,
in which a total luminosity of $5.6\ab^{-1}$ will be accumulated 
at two interaction points for 7 years of running.
There are also other $e^+e^-$ colliders, for instances, 
ILC, FCC-ee and CLIC running at a similar CoM energy.
Here we chose the CEPC as a representative one for
probing the operators considered in this paper.
In our simulations, following kinematical cuts are used to 
estimate the signal significance defined in \eqref{eq:chi2},
\bee
p_{T,\gamma} > 0.5\gev\,, \quad |\eta_{\gamma}| < 2.65\,.
\ene
Fig.~\ref{fig:S:CEPC:O1}, Fig.~\ref{fig:S:CEPC:O2} and Fig.~\ref{fig:S:CEPC:O3} 
show our results for the operators $\calo_{1}$, $\calo_{2}$ and $\calo_{3(4)}$, 
respectively. 

The shaded region in blue represents $L_{X} > 0.1m$ with
assumption $Br(X\to\ell^+\ell^-) = 1$.
As we have mentioned, this condition can be released if the particle $X$
decays into dark particles. For reference, we also show the contour with $L_{X} = 1m$
in dashed-blue line. The gray region in the Fig.~\ref{fig:S:CEPC:O1} is
obtained by data extracted from Ref.~\cite{BaBar:2017tiz},
and stands for a 90\%CL excluded region by the BaBar experiment.
The gray region in Fig.~\ref{fig:S:CEPC:O2} is obtained by reinterpreting
the same data for the operator $\calo_2$, based on the method explained 
in Sec.~\ref{sec:exp:BaBar}.
For the operator $\calo_{3(4)}$, due to suppression of the $s/m_{Z}^2$,
constraint from the BaBar experiment is rather week, and the excluded region
is outside of the plot range in Fig.~\ref{fig:S:CEPC:O3}.
The purple region represents the $1\sigma$ bound of $a_{e}$.
We can see that there is still parameter space, which can account for the 
$a_{e}$, but have not constrained by the existing experiments (if we don't 
consider the astrophysical and cosmological constraints).

Based on the method explained in Sec.~\ref{sec:exp:lep-delphi}
for the DELPHI experiment, the expected exclusion regions at 95\%CL 
are shown in cyan. We can see that for the operator $\calo_1$ the constraint 
in low mass region is slightly weaker than the BaBar limit.
The reasons are, 1) the cross section for the operator $\calo_1$ decreases 
with respect to CoM energy; 2) total luminosity of the DELPHI is about a factor of 3 
larger than the BaBar's one. On the contrary, for the operator $\calo_2$,
DELPHE's constraint is stronger, because on the one hand cross section of the 
signal does not depend on $s$, and on the other hand 
more background events are killed by 
the much central cut on $\cos\theta_\gamma$ (see \eqref{eq:kc:delphe}).
We can see that, inn small region of $m_{X}$, constraint on the Pauli operator 
already reaches to about $60\tev$,
and it is about $2\tev$ for the DMTGC operators.

The black-solid lines show the expected 95\%CL experimental sensitivities at the CEPC
with an integrated luminosity $5.6\ab^{-1}$. The CoM energy of the CEPC is not much
higher than the DELPHE's one, but the luminosity is about 40 times larger,
hence the expected sensitivity enhanced. For low mass senarios, 
the mixing parameter can be probed at a level of $7\times10^{-4}$.
The operator $\calo_2$ with an energy scale $\varLambda_2 \sim 600\tev$ 
can be searched for at the CEPC.
Compared to the sensitivities at future Muon collider~\cite{Casarsa:2021rud},
the CEPC can already probing most of the parameter space.
For the operators $\calo_{3(4)}$, the 95\% sensitivity to $\varLambda_{3(4)}$
can reach to $1\tev$ in the whole mass region (within the plot range),
and it is about $5\tev$ in the low mass region.

\subsection{Significances at ILC}
\label{sec:MA:ES:ILC}
The ILC collider was originally proposed to be run at a CoM energy
$\sqrt{s}=500\gev$~\cite{Bambade:2019fyw},
and recently scenarios with $\sqrt{s}=250\gev$ and 
$\sqrt{s}=1\tev$~\cite{Fujii:2017vwa}
were also considered. Here we focus the the mode with $\sqrt{s}=500\gev$,
at which data will be collected with a total luminosity of $4\ab^{-1}$.
In addition, the H20 running scenario~\cite{Barklow:2015tja}, 
in which both electron and position beams are polarized, was aiming to
optimize the physics performance of the experiment.
In this paper we consider three polarization configurations, which are listed in
the Tab.~\ref{tab:ILC:Pol}.
\begin{table}[t]
\renewcommand{\arraystretch}{1.3}
\setlength\tabcolsep{6pt}
\caption{Polarization configures and the corresponding luminosities studied in this paper.}
\begin{center}
\begin{tabular}{c|ccc}
\hline 
$(P_{e^{-}},  P_{e^{+}}) $ & (0,\,0) & (+80\%,\,0)  & (+80\%,\,-30\%)  
\\ \hline
$\call_{\rm Int.}$ $\left[ \ab^{-1} \right] $ & $4$ & $1.6$  & $1.6$  
\\ \hline
\end{tabular}
\end{center}
\label{tab:ILC:Pol}
\end{table}
The experimental significances are estimated by applying following 
kinematical cuts~\cite{Habermehl:2020njb},
\bee
p_{T,\gamma} > 6\gev\,, \quad |\eta_{\gamma}| < 2.79\,.
\ene
\begin{figure}[ht]
\begin{center}
\subfigure[]{
\label{fig:S:ILC:O1}
{\includegraphics[width=0.3\linewidth]{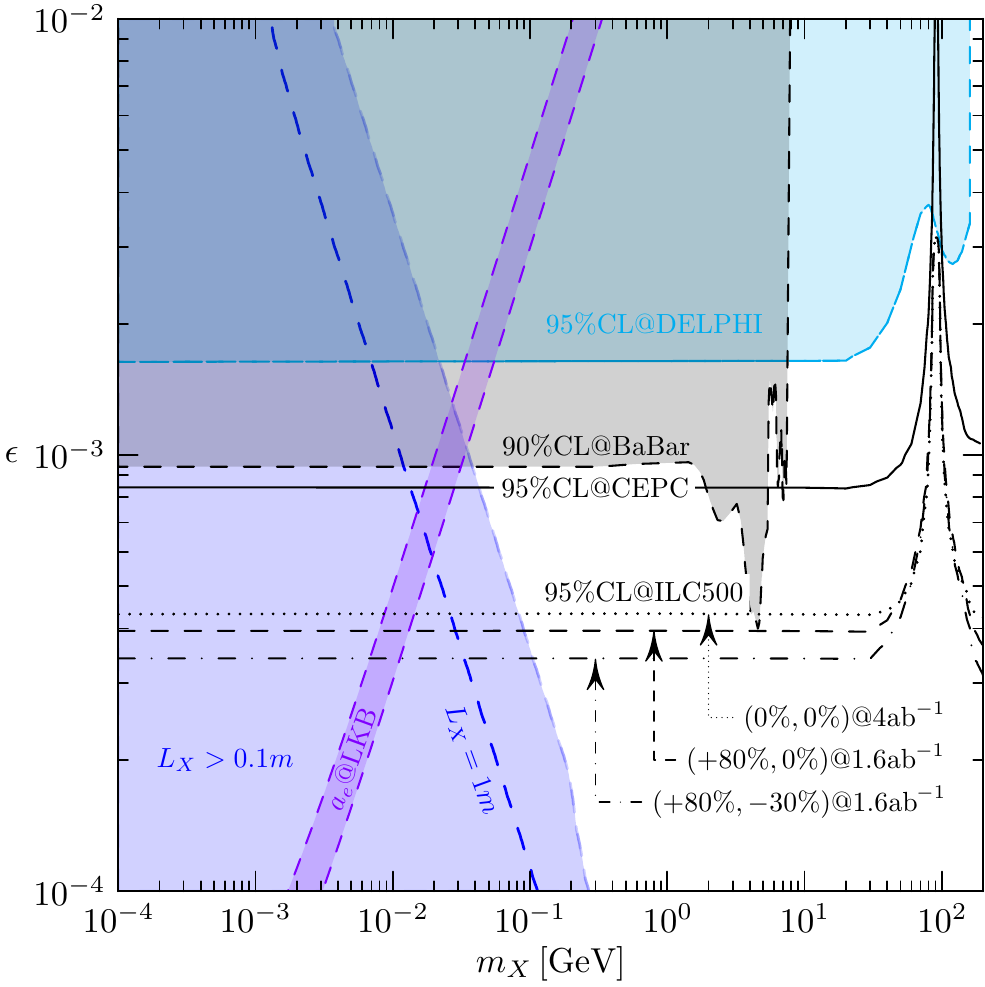}}
}
\subfigure[]{
\label{fig:S:ILC:O2}
{\includegraphics[width=0.3\linewidth]{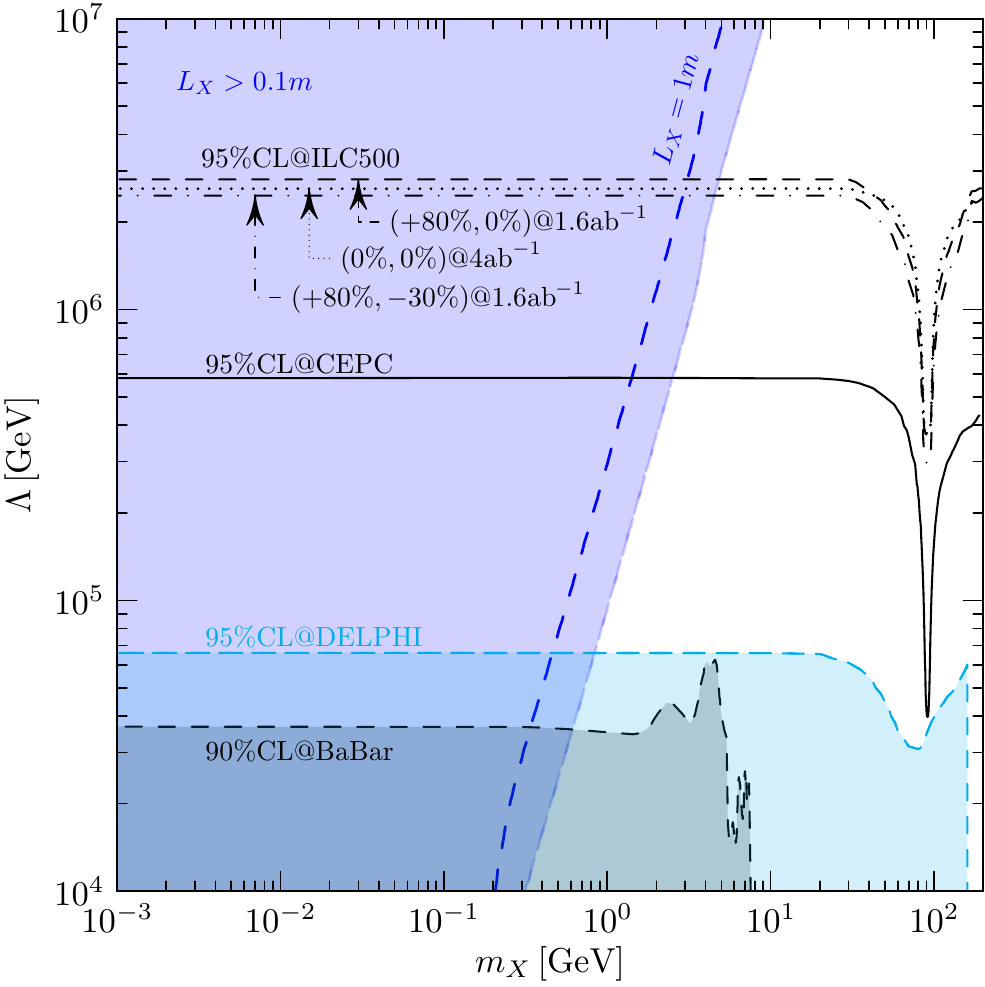}}
}
\subfigure[]{
\label{fig:S:ILC:O3}
{\includegraphics[width=0.3\linewidth]{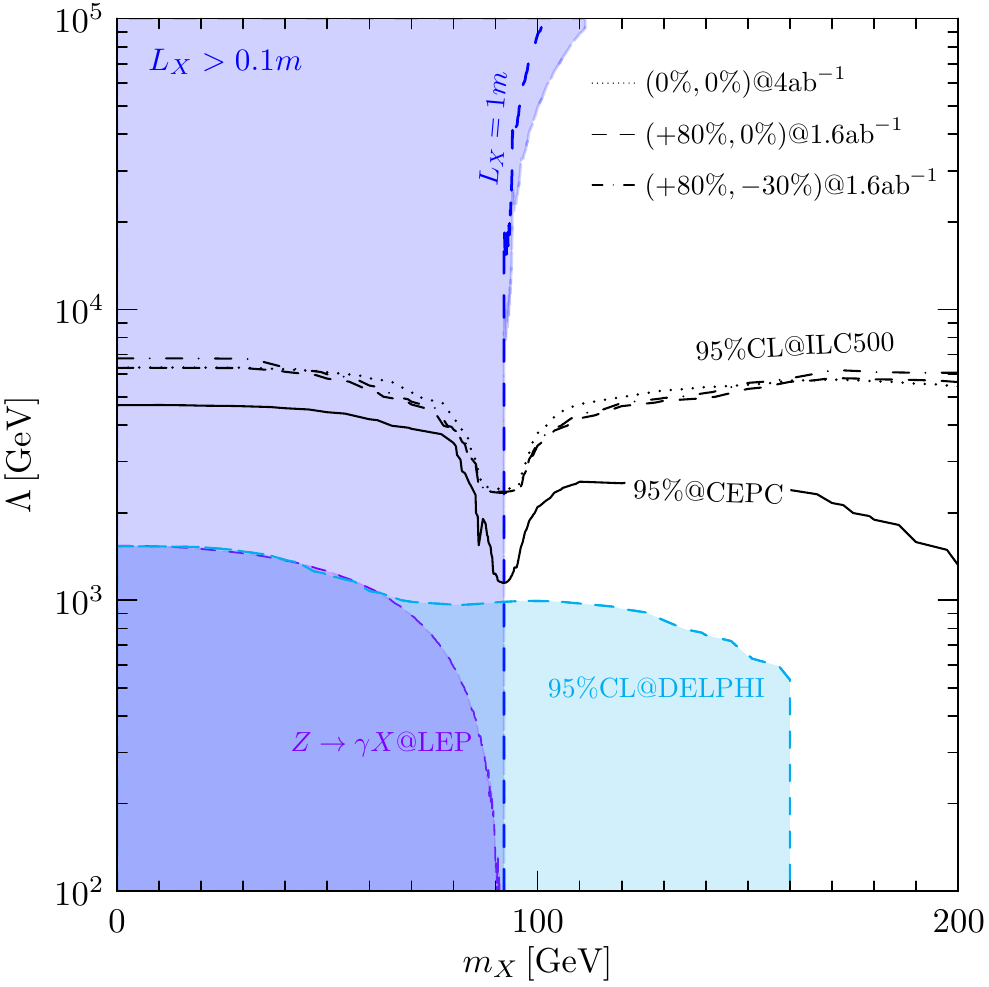}}
}
\caption{\it
Expected sensitivities at  95\%CL for the operators 
$\calo_1$ (Fig.~\ref{fig:S:ILC:O1}), $\calo_2$(Fig.~\ref{fig:S:ILC:O2}) 
and $\calo_{3(4)}$ (Fig.~\ref{fig:S:ILC:O3}) at the ILC500.
The shades regions are the same as 
the ones shown in Fig.~\ref{fig:S:CEPC}.
The black-dotted, -dashed, -dashdotted lines show cases for
$\big(P_{e^-}, P_{e^+}\big) = (0\%, 0\%), (+80\%, 0\%)$, 
and $(+80\%, -30\%)$, respectively.
The 95\%CL  line for the CEPC is also show in black-solid line 
for reference.
}
\label{fig:S:ILC}
\end{center}
\end{figure}
Fig.~\ref{fig:S:ILC:O1}, Fig.~\ref{fig:S:ILC:O2} and Fig.~\ref{fig:S:ILC:O3}
show the 95\%CL sensitivities for the operators $\calo_1$, $\calo_2$ and
$\calo_{3(4)}$, respectively. The shades regions are the same as 
the ones shown in Fig.~\ref{fig:S:CEPC:O1}-\ref{fig:S:CEPC:O3}.
For the the operator $\calo_1$ we can see that, 
even through the total luminosity of the ILC is smaller,
and the cross section of the signal is reduced,
the experimental sensitivity is
enhanced by roughly a factor of 2 for $\big(P_{e^-}, P_{e^+}\big) = (0\%, 0\%)$.
This is due to 1) in low energy region the background also
decreases with respect to $s$ as shown in Fig.~\ref{fig:MA:XE-YXS};
2) the stronger transverse momentum cut kills more background.
However, since the background tends to be a constant at higher $s$,
such enhancement is not expected at colliders with higher CoM energy.
On the other hand, as expected the polarization configuration
$\big(P_{e^-}, P_{e^+}\big) = (+80\%, -30\%)$ gives better sensitivity.
But the enhancement is not so promising, since the projected total luminosity
is $1.6\ab^{-1}$ which is smaller than the one of unpolarized scenario 
by a factor of more than 2.
Because of the same reason, polarized beams can not provide 
sizable optimization for the operators $\calo_2$ and $\calo_{3(4)}$,
as shown in Fig.~\ref{fig:S:ILC:O2} and Fig.~\ref{fig:S:ILC:O3}, respectively.
On the other hand, sensitivity to the operator  $\calo_2$ is significantly 
enhanced at the ILC500, and reaches to a level of $\sim10^3\tev$ which
is much higher than the expectation at the future Muon collider~\cite{Casarsa:2021rud}.
This is due to the 
distinctive polar angle distribution between the signal and background, 
as shown in Fig.~\ref{fig:DXS:cosa}, and hence
the background is strongly suppressed by the stronger cut $p_{T,\gamma} > 6\gev$.
Similar enhancement happens for the operator $\calo_{3(4)}$,
but it is reduced because the cross section depends on $\varLambda^{-4}_{3(4)}$.
On the other hand, probing power in the high mass region is significantly enhanced.
The sensitivity at 95\%CL can reach to a level of $7\tev$, which is 
about 5 times larger then the exclusion limit given by $Z\to\gamma X$ at the LEP
and $e^{+}e^{-} \to \gamma X$ by the DELPHE in the low mass region.

\section{Conclusion}
\label{sec:conclusion}
In summary we studied mono-$\gamma$ production induced by photon-like,
Pauli and DMTGC operators at future $e^+e^-$ colliders.
We show that, while energy of the photon is purely determined by 
kinematics, but polar angle distribution of the photon are very distinctive
for the various operators and the background. Particularly, for the DMTGCs,
$X$ and $\gamma$ are generated via a $s$-channel virtual $Z$-boson,
hence photons are dominantly produced in the central region,
as shown in Fig.~\ref{fig:DXS:cosa}.
Furthermore, behaviors of the total cross sections at high energy region
are also different. While contribution of the photon-like operator decreases as $1/s$,
and the one of the Pauli operator does not depend on $s$,
cross sections for the DMTGCs grow quickly with respect to
$s$. Hence at high energy colliders the DMTGC operators are more promising 
comparing to the Pauli operators.

There were already $e^+e^-$ experiments that searched for dark particles
via the mono-$\gamma$ channel. Focus on the BarBar~\cite{BaBar:2017tiz}
and DELPHI~\cite{DELPHI:2008uka} experiments,
we re-interpret the results as constraints on the parameters considered in
this work. In small mass region of $m_{X}$, constraint from the DELPHI
experiment on the Pauli operator already reaches to about $60\tev$,
and it is about $2\tev$ for the DMTGC operators.
We also considered the anomalous magnetic dipole moment
of the electron~\cite{Morel:2020dww,Darme:2020sjf}.
We show that there is still parameter space, which can account for the 
$a_{e}$, but have not constrained by the existing experiments (except for 
astrophysical and cosmological constraints).
We further studied the expected experimental significance at the CEPC
and the ILC. Our results indicate that very high energy colliders, 
for instance Muon colliders at $3, 10\tev$, can not
give much more profound limit on the Pauli operators.
This is because cross section for the Pauli operator does not depend on $s$.
The 95\% lower limit on $\varLambda_2$ can reach to $600\tev$ at the CEPC,
and can be enhanced to $\sim 10^3\tev$ at ILC500.
On the other hand, because production rates induced by the DMTGC operators
grow quickly as increasing $s$, it is more interesting to search for signals
of the DMTGC at high energy colliders.
The expected lower limit of $\varLambda_{3(4)}$ at 95\%CL is about $5\tev$
at the CEPC, and it is about $7\tev$ at the ILC500.
Possible enhancement by beam polarization are also studied.
However, because of its smaller integrated luminosity, 
the bounds on the 
scale parameters can only be enhanced slightly.


\acknowledgments
K.M. was supported by the Innovation Capability Support Program of Shaanxi
Province (Program No. 2021KJXX-47).

\bibliographystyle{mArticle}
\bibliography{mArticle}

\end{document}